\documentclass[10pt,aps,prd,nofootinbib,superscriptaddress,
  twocolumn,preprintnumbers,balancelastpage]{revtex4}
\PassOptionsToPackage{slash}{url}
\usepackage{graphicx,epsfig,psfrag,amssymb,hyperref}
\usepackage{multirow}
\usepackage{color,graphicx,epsfig,psfrag,amsmath,empheq}
\usepackage{bm}
\usepackage{mathrsfs,amsfonts,color}
\usepackage{slashed}
\usepackage{hepunits}
\usepackage{color}
\usepackage{enumitem}
\usepackage{booktabs}
\usepackage{xspace}
\usepackage[utf8]{inputenc}

\AtBeginDocument{
\heavyrulewidth=.08em
\lightrulewidth=.05em
\cmidrulewidth=.03em
\belowrulesep=.65ex
\belowbottomsep=0pt
\aboverulesep=.4ex
\abovetopsep=0pt
\cmidrulesep=\doublerulesep
\cmidrulekern=.5em
\defaultaddspace=.5em
}

\newcommand{\cP}{\mathcal{P}}
\newcommand{\cL}{\mathcal{L}}
\newcommand{\br}{\ensuremath{\mathrm{BR}}\xspace}
\newcommand{\abs}[1]{\left\lvert#1\right\rvert}
\newcommand{\mb}{\ensuremath{\mathrm{mb}}\xspace}
\newcommand{\pb}{\ensuremath{\mathrm{pb}}\xspace}
\newcommand{\ifb}{\ensuremath{\mathrm{fb}^{-1}}\xspace}
\newcommand{\pt}{\ensuremath{p_\mathrm{T}\xspace}}
\newcommand{\track}{\ensuremath{\mathrm{trk}}\xspace}
\newcommand{\doca}{\ensuremath{\mathrm{DOCA}}\xspace}
\newcommand{\ip}{\ensuremath{\mathrm{IP_{T}}}\xspace}
\renewcommand{\TeV}{\ensuremath{\mathrm{Te\mkern-1.5muV}}\xspace}
\renewcommand{\GeV}{\ensuremath{\mathrm{Ge\mkern-1.5muV}}\xspace}
\renewcommand{\MeV}{\ensuremath{\mathrm{Me\mkern-1.5muV}}\xspace}
\renewcommand{\keV}{\ensuremath{\mathrm{ke\mkern-1.5muV}}\xspace}
\renewcommand{\sec}{\ensuremath{\mathrm{sec}}\xspace}
\newcommand{\TM}{\ensuremath{\mathcal{T\!M}}\xspace}
\newcommand{\sig}{\ensuremath{\sigma_\mathrm{stat}}\xspace}
\newcommand{\res}{\sigma_{m_{ee}}}
\newcommand{\resg}{\sigma_{m_{ee\gamma}}}

\begin{document}

\title{Discovering True Muonium at LHCb}

\author{Xabier Cid Vidal}
\email{xabier.cid.vidal@cern.ch}
\affiliation{Instituto Galego de Fısica de Altas Enerxıas (IGFAE),
  Universidade de Santiago de Compostela, Santiago de Compostela,
  Spain}

\author{Philip Ilten}
\email{philten@cern.ch}
\affiliation{School of Physics and Astronomy, University of
  Birmingham, Birmingham, B152 2TT, UK}

\author{Jonathan Plews}
\email{jonathan.plews@cern.ch}
\affiliation{School of Physics and Astronomy, University of
  Birmingham, Birmingham, B152 2TT, UK}

\author{Brian Shuve}
\email{bshuve@g.hmc.edu}
\affiliation{Harvey Mudd College, 301 Platt Blvd., Claremont, CA
  91711, USA}
\affiliation{University of California, 900 University Ave., Riverside,
  CA 92521, USA}

\author{Yotam Soreq}
\email{yotam.soreq@cern.ch}
\affiliation{Theoretical Physics Department, CERN, CH-1211 Geneva 23,
  Switzerland}
\affiliation{Department of Physics, Technion, Haifa 32000, Israel}

\begin{abstract}
We study the potential of the LHCb experiment to discover, for the
first time, the $\mu^+\mu^-$ true muonium bound state. We propose a
search for the vector $1^3S_1$ state, $\TM$, which kinetically mixes
with the photon and dominantly decays to $e^+e^-$. We demonstrate
that a search for $\eta \to \gamma\TM$, $\TM\to e^+e^-$ in a displaced
vertex can exceed a significance of 5 standard deviations assuming
statistical uncertainties. We present two possible searches: an
inclusive search for the $e^+e^-$ vertex, and an exclusive search
which requires an additional photon and a reconstruction of the $\eta$
mass.
\end{abstract}

\preprint{CERN-TH-2019-049}
\maketitle

\section{Introduction}
\label{sec:intro}
Electromagnetic~(EM) interactions between oppositely charged particles
form bound states; by far, the most well known of these are the
atoms. Similar atom-like bound states of elementary particles have
since been discovered, including positronium (a bound state of
$e^+e^-$)~\cite{Deutsch:1951zza} and muonium (a bound state of
$\mu^+e^-$)~\cite{Hughes:1960zz}. The properties of these bound states
are predicted by quantum electrodynamics~(QED), and measurements of
the mass and spectra provide precision tests of QED.

However, there remain heavier QED bound states that have not yet been
experimentally observed which can provide unique probes that are
sensitive to beyond the standard model (BSM) physics.  In particular,
the hypothesized bound state known as true muonium
$(\mu^+\mu^-)$~\cite{Hughes:1971} has yet to be discovered. In this
work, we explore the potential of the LHCb experiment to discover the
lowest spin-1 state of true muonium via its displaced decays to
$e^+e^-$ pairs. We show that true muonium can be observed with a
statistical significance exceeding $5$ standard deviations using the
expected 15\,\ifb of LHC Run~3 data to be collected with the upgraded
LHCb detector~\cite{Bediaga:2012uyd,Bediaga:2013bkh,
  Bediaga:2013iwy,Bediaga:2014vzo,Bediaga:2014tuj,Bediaga:2018lhg}.

The most promising true muonium state for discovery is the $1^3S_1$
state, which in the non-relativistic limit has zero orbital angular
momentum and is in the spin-triplet state. This vector muonium state,
which we denote as $\TM$, kinetically mixes with the photon resulting
in a phenomenology similar to the dark
photon~\cite{Okun:1982xi,Galison:1983pa,Holdom:1985ag,Pospelov:2007mp,
  ArkaniHamed:2008qn,Bjorken:2009mm}. Dark photons have been the
subject of much recent study,
\textit{e.g.}~\cite{Alexander:2016aln,Beacham:2019nyx,Essig:2013lka},
allowing us to use these latest developments in the discovery of \TM
at LHCb. Note that spin-singlet true muonium states also exist, but
their dominant decay are to $\gamma\gamma$, which is challenging to
reconstruct with the LHCb detector. Therefore, we concentrate on the
discovery of \TM, the spin-triplet true muonium state.

Other possible search avenues for \TM are with the currently running
HPS experiment~\cite{Baltzell:2016eee} or via rare $B$ decays into
leptonium at LHCb~\cite{Fael:2018ktm}. However, both of these methods
are statistically limited with potentially large backgrounds and are
not expected to have discovery potential. The proposed
RedTop~\cite{Gatto:2016rae,Ji:2018dwx} experiment at Fermilab is
designed to produce a large flux of $\eta$ mesons, and using the
methods outlined in this work, might also be sensitive to \TM.
Searching for a $\TM\, \gamma$ final state from $e^+ e^-$ collisions
has also been proposed~\cite{Brodsky:2009gx}, which may be accessible
to Belle~II. However, \TM discovery is not expected given the Belle~II
dark photon reach~\cite{Kou:2018nap}.

The rest of the paper is organized as follows.  In
Sections~\ref{sec:TMDP} and~\ref{sec:dissociation} we describe the
analogy between \TM and dark-photon and highlight the differences.
Section~\ref{sec:LHCbSearch} contains the details of the proposed LHCb
search.  We conclude in Section~\ref{sec:outlook}.  The appendices
contains technical details and a discussion about \TM and new physics.

\section{True Muonium Signal as a Dark Photon}
\label{sec:TMDP}
Dark photons are massive spin-1 states that couple via a kinetic
mixing $\varepsilon$ to the standard model~(SM) photon:

\begin{align}
	\label{eq:kinmix}
  	\mathcal{L} \supset \frac{\varepsilon}{2}\,F_{\mu\nu}F'^{\mu\nu} \,,
\end{align}

where $F^{\mu\nu}$ and $F'^{\mu\nu}$ are the dark photon and SM photon field strengths, respectively. 
The phenomenology of $\TM$ is similar to that of a dark photon, and the mass and kinetic mixing are predicted by QED at leading order:

\begin{align}
  	\label{eq:mTM}
  	& m_{\TM} = 2m_\mu - B_E \approx 211\,\MeV\,, \\
  	\label{eq:epsTM}
  	& \varepsilon_{\TM} = \alpha^2/2 \approx 2.66 \times 10^{-5} \, ,
\end{align}

where $B_E \approx m_\mu \alpha^2/4=1.41\,\keV$ is the $\TM$ binding
energy, estimated in the non-relativistic limit.  Our result is in
agreement with Ref.~\cite{An:2015pva}, where the kinetic mixing of
hidden sector onium states was calculated.  We emphasise that the
above analogy between \TM and the dark photon is valid only at
energies close to the \TM mass, as relevant to our study.

As noted earlier, $\TM$ decays through the same kinetic mixing to an
$e^+e^-$ final state with a branching fraction of $\br( \TM\to
e^+e^-)\approx 98\,$\%, while the sub-dominant decay mode has $\br(
\TM\to 3\gamma)\approx 1.7\,$\%\,.  The $\TM$ lifetime at leading
order is

\begin{align}
  \label{eq:tauTM}
  \tau_{\TM} \approx \frac{6}{\alpha^5 m_\mu} \approx 1.8 \times
  10^{-12} \, \sec \,.
\end{align}

Because of the forward coverage of LHCb, light particles produced
within LHCb acceptance typically have large boosts. Given the expected
boost of \TM within LHCb and the relatively long proper lifetime of
$0.53\,$mm, the decay of \TM into $e^+ e^-$ within LHCb will typically
produce a resolvable displaced vertex. While searches for long-lived
particles typically focus on new BSM states~\cite{Alimena:2019zri},
\TM is an example of a SM long-lived particle that can be searched for
at LHCb. Predictions of the mass and lifetime at higher order than
those derived here are available~\cite{Lamm:2017ioi,Ji:2018dwx};
however, it is unlikely that LHCb will be sensitive to these higher
order corrections.

Since \TM and dark photon phenomenology are similar, excluding \TM
dissociation detailed in Section~\ref{sec:dissociation}, projected
dark photon reaches from future experiments can provide a rough guide
to \TM sensitivity. In Fig.~\ref{fig:aprime} the dark photon parameter
space is plotted in dark photon mass ($m$) and kinetic mixing
($\varepsilon$) using \textsc{Darkcast}~\cite{Ilten:2018crw}, where
\TM corresponds to a single point given by the $\varepsilon$ and $m$
of Eqs.~\eqref{eq:mTM} and~\eqref{eq:epsTM}. The gray regions
correspond to already excluded parameter space, while the colored
regions represent possible reach from relevant future
experiments. Dashed lines indicate experiments where dissociation will
be an issue. These include searches by FASER~\cite{Ariga:2018uku},
SeaQuest~\cite{Gardner:2015wea}, and SHiP~\cite{Alekhin:2015byh} where
\TM will dissociate as it passes through the shielding.

Both the proposed LHCb $D^{*0} \to D^0 A'(\to e^+
e)$~\cite{Ilten:2015hya} and inclusive $A'(\to \mu^+
\mu^-)$~\cite{Ilten:2016tkc} searches are shown, to demonstrate how
dark photon searches based on this study could be used to fill the gap
between the two searches. The dashed regions for these LHCb searches
correspond to post-module search strategies where the \TM will
dissociate. The expected displaced reach of
HPS~\cite{Baltzell:2016eee} does not cover the \TM parameter space
point, and will also suffer from some dissociation. Additionally, the
expected prompt Belle~II reach~\cite{Kou:2018nap} does not extend to
large enough lifetimes to discover \TM, and the nominal Belle II
lifetime resolution will not be sufficient for effective displaced
searches.

\begin{figure}
  \includegraphics[width=0.49\textwidth]{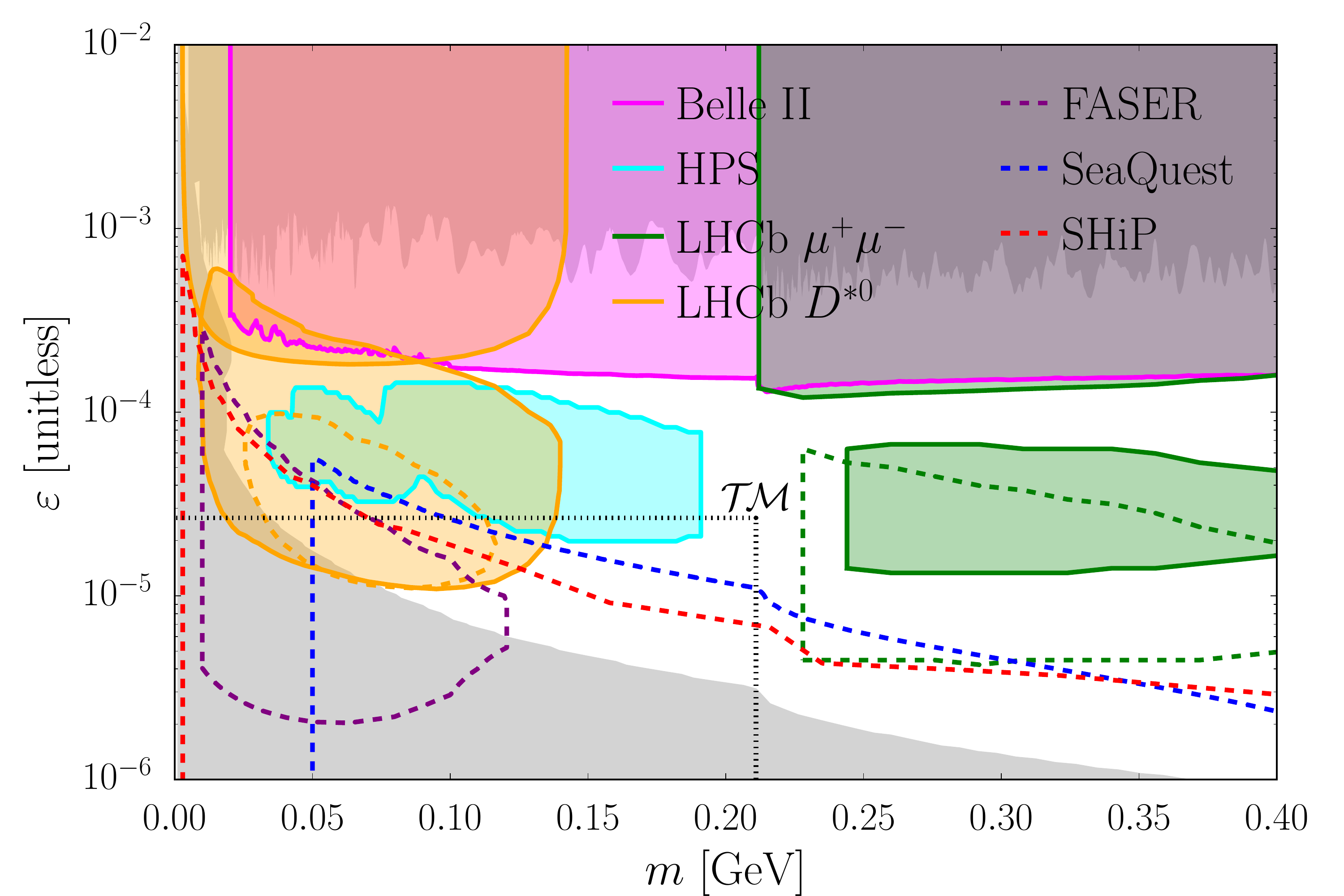}
  \caption{Dark photon parameter space in dark photon mass and kinetic
    mixing with (gray) previous limits and future reach from (magenta)
    Belle II, (purple) FASER, (cyan) HPS, and (green/yellow) LHCb. \TM
    corresponds to the marked point, using Eqs.~\eqref{eq:mTM} and
    \eqref{eq:epsTM}.
    \label{fig:aprime}}
\end{figure}

\section{Dissociation of True Muonium}
\label{sec:dissociation}
Because $\TM$ is a bound state rather than an elementary particle,
there are significant differences between \TM and dark photon
phenomenology. Most importantly, $\TM$ can dissociate when the
constituent muons of the bound state interact with the detector
material, resulting in a separated $\mu^+$ and $\mu^-$ with an
invariant mass just above the mass of \TM, $m_{\TM}$.

The $\TM$ dissociation cross section is estimated to
be~\cite{Mrowczynski:1985qt,Holvik:1986ty,Mrowczynski:1987gq,Denisenko:1987gr}
$\sigma_{\TM \to \mu\mu}\approx 13 Z^2\,\mathrm{b}\,$, where $Z$ is
the atomic number of the material inducing the dissociation. The bulk
of the material traversed by \TM within LHCb prior to its decay is the
aluminum radio frequencey (RF)-foil (made of AlMg3) and the silicon
vertex locator (VELO) sensors.  Since both aluminum and silicon have
similar $Z$ and number densities, the mean free path for \TM
traversing the material of the detector is,

\begin{align}
  \lambda^{-1} = \sigma_{\TM \to \mu\mu} n_a \approx 13\,\mm^{-1} \, ,
\end{align}

where the number density is $n_a\approx
6.0\,(5.0)\times10^{19}\,\mathrm{atoms}/\mm^3$~\cite{Tanabashi:2018oca}
and $Z=13\,(14)$ for aluminum\,(silicon). Thus, the probability of \TM
dissociating is given by

\begin{equation}
  \cP_\mathrm{dis} = 1 - e^{-x/\lambda} \, ,
\end{equation}

where $x$ is the distance of the material traversed. The RF-foil will
have a nominal width of $0.25\,\mm$ in Run~3 and the VELO sensors a
nominal width of $0.2\,\mm$. Consequently, every encounter of \TM with
material in the VELO results in a minimum dissociation probability of
$\cP_\mathrm{dis}\gtrsim 90\%$.

Given the expected material budget of the LHCb detector during
Run~3~\cite{Bediaga:2013bkh}, the boost distribution for \TM produced
within LHCb acceptance, and $\cP_\mathrm{dis}$, roughly half of the
\TM produced are expected to dissociate without decaying into an
$e^+e^-$ final state. The radial flight distance distribution of the
\TM particles which do decay into $e^+ e^-$, is compared to the
expected $e^+ e^-$ background in Fig.~\ref{fig:radial}. On average,
\TM has a higher boost than the background, resulting in a flatter
distribution that is abruptly truncated by dissociation.

This dissociation gives rise to a signal of $\mu^+\mu^-$ originating
from the regions of high material density at LHCb. While nearly half
of the \TM produced is a considerable fraction of the total signal,
the dissociated $\mu^+\mu^-$ signal is difficult to reconstruct and
suffers from large irreducible backgrounds. The two muons will be
nearly collinear and will typically share hits within the VELO,
resulting in poorly defined tracks. Additionally, since the
dissociation occurs in material, the conversion background of $\gamma
\to \mu^+ \mu^-$ can no longer be eliminated with a material veto
without eliminating the signal itself. Therefore, for the remainder of
the paper we focus on the 50\% of signal events which decay via
$\TM\rightarrow e^+e^-$.

\begin{figure}[t]
\includegraphics[width=0.5\textwidth]{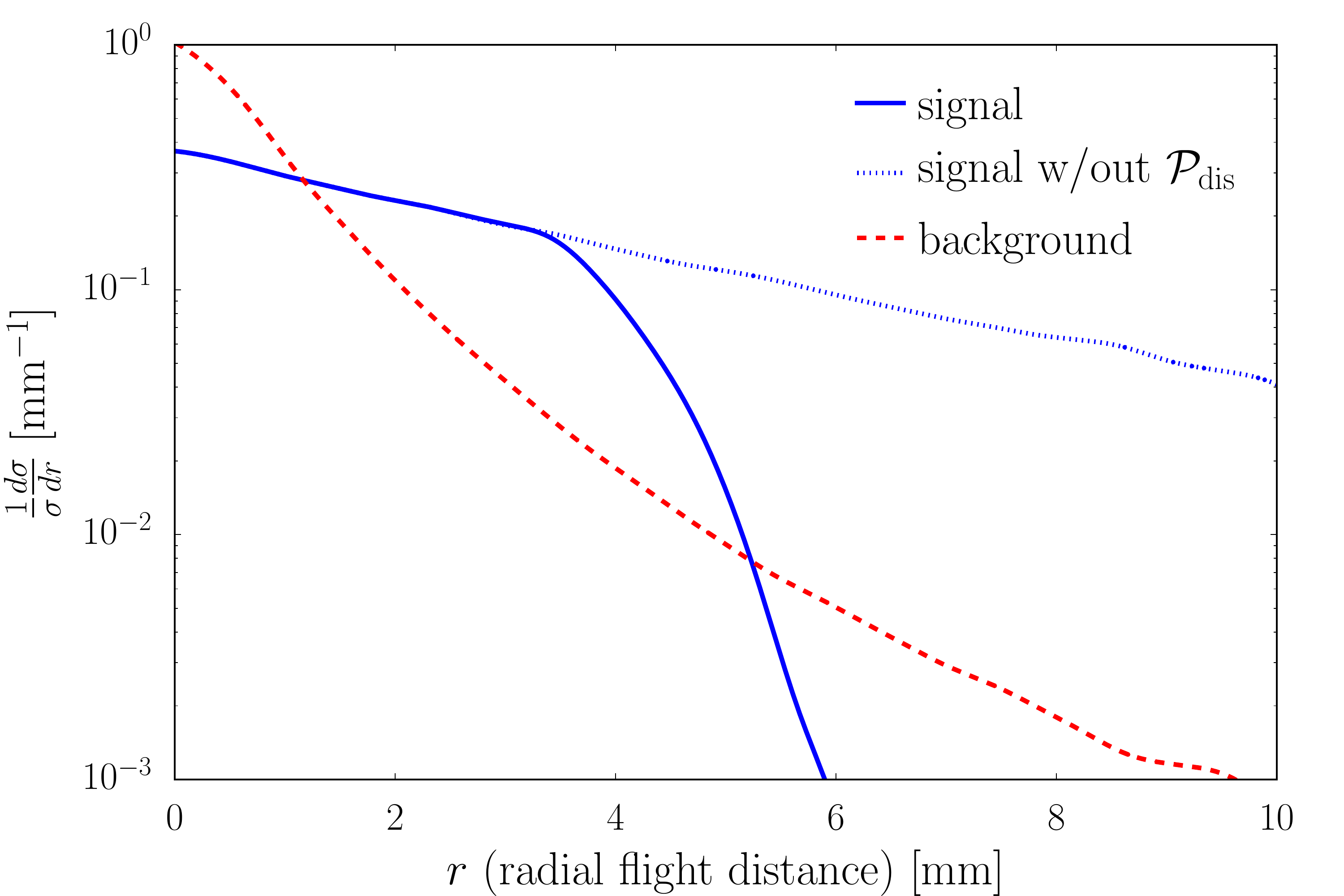}
\caption{Normalized radial flight distance distributions for the $\TM
  \to e^+ e^-$ signal (blue solid) with dissociation, (blue dotted)
  without dissociation, and (red dashed) the $e^+ e^-$ background from
  $B$-hadron decays.
  \label{fig:radial}}
\end{figure}

\section{Proposed LHCb Search}
\label{sec:LHCbSearch}
We propose searching for $\TM$ as a displaced $e^+e^-$ resonance.
Since $\TM$ behaves like a dark photon, the signal rate can be
calculated directly from the off-shell photon rate as given by the
prompt $e^+e^-$ spectrum
data~\cite{Bjorken:2009mm,Ilten:2016tkc,Aaij:2017rft}. For any initial
($Y$) and final ($X$) states, the ratio between the number of $Y \to
X\, \TM \to X\, e^+e^-$ events, $S_{\TM}$, and the number of prompt
$e^+e^-$ events, $Y \to X \gamma^* \to X e^+e^-$, $B_\mathrm{EM}$, is
fixed. For the $e^+e^-$ invariant mass within the range of
$|m_{ee}-m_{\TM}|<2\res$, where $\res$ is the $e^+e^-$ invariant mass
resolution, this ratio is given by

\begin{align}
  \label{eq:SB}
  \frac{S_{\TM}}{B_\mathrm{EM}} \approx \frac{3\pi}{16}\frac{m_\mu}{\res}
  \alpha^3 \approx \frac{20\,\MeV}{\res}\, 1.2 \times 10^{-6} \, .
\end{align}

The dominant source of off-shell photons in the mass range
$m_{\TM}\approx211\,\MeV$ is from $\eta \to \gamma \gamma^*$
decays. We therefore focus on searching for $\TM$ produced from $\eta
\to \gamma\, \TM$ decays with a $\TM\to e^+e^-$ final state. The
signal can be fully normalized by the data using the procedure
outlined above. The ratio of Eq.~\eqref{eq:SB} must be corrected by
the different acceptance and efficiency factors for a displaced
$e^+e^-$ signal relative to the prompt signal. Additionally, the
signal rate should be corrected by the expected dissociation factor,
to account for $\TM$ that dissociate without decaying.

The number of signal events can be estimated as follows:~we simulate
in \textsc{Pythia\,8.2}~\cite{Sjostrand:2014zea} both the $pp$ total
cross section, $\sigma_\mathrm{tot}=100\,\mb$, and the average number
of $\eta$ mesons produced per collision within the LHCb acceptance,
$N_\eta = 0.83$. The former is in agreement with the LHCb inelastic
cross-section measurement~\cite{Aaij:2018okq}, while the latter
correctly predicts the low mass limit of the LHCb inclusive
$\mu^+\mu^-$ dark photon search~\cite{Aaij:2017rft}. Given that
$\br(\eta\to\gamma\,\TM)=4.8\times 10^{-10}$~\cite{Ji:2018dwx}, which
agrees well with Eq.~\eqref{eq:SB} using the differential $\eta \to
\gamma e^+ e^-$ shape from \textsc{Pythia}, the signal cross section
in the fiducial volume is

\begin{align}
  \label{eq:TMXsec}
  \sigma^\mathrm{fid}_{\TM} = \sigma_\mathrm{tot} N_\eta \,
  \br(\eta\to\gamma\,\TM) \approx 40\, \pb \, .
\end{align}

In our analysis we consider two possible search strategies:
(i)~inclusive search -- the final state is $e^+e^-$ and we do not
search for the photon, thus the $\eta$ is not reconstructed (in
principle this search is sensitive to any \TM production mechanism);
(ii)~exclusive search -- the final state is $\gamma\,e^+e^-$ and the
$\eta$ is reconstructed. Each of these methods has both advantages and
disadvantages. The inclusive search is simpler and expected to have
smaller systematic uncertainties, while the background rates for the
exclusive analysis are smaller. Without a full detector simulation and
data-driven background estimates with their corresponding
uncertainties, we cannot definitively state which of the two
strategies is optimal; we therefore estimate the potential
sensitivities of both. The details of our signal and background
simulations are provided in Appendix \ref{app:simulation_details}.

The LHCb experiment is a forward arm spectrometer which covers
pseudorapidities between 2 and
5~\cite{Alves:2008zz,Aaij:2014jba}. This is a simplification of the
coverage provided by the individual sub-systems, but provides an
adequate description, given the evolving nature of the upgraded
detector and the weak assumptions made on electron identification
efficiencies in this paper.  While the exact performance of LHCb
during Run~3 and~4 is yet to be fully understood, we estimate the
relevant quantities as follows, with more details given in
Appendix~\ref{sec:LHCbPref}.  The $e^+e^-$ invariant mass resolution
around the \TM mass is estimated to be $\res\approx20\,\MeV$, based on
the $K_S^0 \rightarrow e^+e^-e^+e^-$ LHCb
study~\cite{MarinBenito:2193358}, while $\resg$ around the $\eta$ mass
is estimated to be $50\,\MeV$ based on
Refs.~\cite{Aaij:2014jba,LHCb:2012af,Aaij:2016ofv}.

We apply the following baseline selection criteria for both cases (i)
and (ii):

\begin{enumerate}
\item Two opposite-sign electrons in the LHCb acceptance and with
  $p(e^\pm)>10\,\GeV$, $p_T(e^\pm)>0.5\,\GeV$, and transverse impact
  parameter~(\ip) which is not consistent with zero, $\ip(e^\pm) >
  3\sigma_{\ip(e)}$, where $\sigma_{\ip(e)}$ is the \ip resolution;
\item A reconstructed $\TM\!\!\to\!\! e^+ e^-$ candidate in the LHCb acceptance and with
  $\pt(\TM)\!>\!1.0\,\GeV$, $\abs{m_{ee} -
  m_{\TM}} < 2\res$, and the distance of closest approach~(\doca)
  between the two electrons consistent with zero, $\doca(e^+,e^-) <
  3\,\sigma_{\doca(e^+,e^-)}$ (the details on \doca resolution are
  given in the Appendix~\ref{sec:LHCbPref}). This ensures that the electron
  pair forms a high-quality vertex.
\end{enumerate}

For case~(ii), in which we reconstruct the additional photon from the
$\eta$ decay, there are two additional baseline selections:

\begin{enumerate}[resume]
\item A photon in the LHCb acceptance and
$p(\gamma)>5\,\GeV$, and $\pt(\gamma) > 0.65\,\GeV$;

\item A reconstructed $\eta$ candidate within the LHCb acceptance and
  $\abs{m_{ee\gamma} - m_\eta} < 2\resg$.
\end{enumerate}

For both cases (i) and (ii), data is expected to be collected using an
$e^+ e^-$ trigger. During Run~1 and~2, only a single electron trigger
with tight kinematic cuts was available in the first-level hardware
trigger, which is not efficient for this signal. However, in Run~3
and~4 full online reconstruction with triggerless readout will be
available \cite{Bediaga:2014vzo}, which will allow the reconstruction
of lower momentum signals such as the electrons from $\TM$
decays. Because $\TM$ decays are displaced and inside a narrow
invariant mass window, the $\TM$ candidates can be reconstructed and
recorded in Run~3 and~4 with a high efficiency.

The dominant background after the baseline selection is from
$B$-hadron decays, which are also displaced. Decays of $D$-hadrons are
a sub-dominant background since these rarely produce an $e^+e^-$ pair
which creates a reconstructible vertex in the chosen kinematic
regime. The background from photon conversions was also estimated and
found to be sub-dominant, using techniques from the proposed $D^{*0}
\to D^0 e^+ e^-$ dark photon search~\cite{Ilten:2015hya} and a
material veto similar to that used in the LHCb inclusive $\mu^+ \mu^-$
dark photon search~\cite{Alexander:2018png}. In the same regard, the
background from $\eta \rightarrow e^+e^- \gamma$ decays will be also
sub-dominant taking into account the expected displacement of the \TM
before decaying (see Fig.~\ref{fig:radial}). Given the excellent LHCb
resolution for reconstructing the signal decay vertex
\cite{LHCbVELOGroup:2014uea}, a moderate cut in this displacement
would be enough to reduce this background to negligible levels.

$B$-mesons tend to decay to a high multiplicity of tracks that
originate from the same decay vertex. These events are, in principle,
readily suppressed by $B$-decay vetoes used in the LHCb dark photon
search~\cite{Aaij:2017rft} and $B_s^0 \to \mu^+\mu^-$ lifetime
measurement~\cite{Aaij:2017vad}. As a simple proxy for these vetoes,
we apply the following additional selections:

\begin{enumerate}[resume]
\item The \TM candidate is isolated from other tracks in the LHCb
  acceptance:~tracks with $\pt(\track) > 0.5\,\GeV$ and $\ip(\track) >
  3\sigma_{\ip(\track)}$ must satisfy $\doca(\track,e) >
  3\sigma_{\doca(\track,e)}$ for both electrons.
\item The opening angle, $\theta$, between the flight and momentum
  vectors of the \TM candidate is consistent with zero. The resolution
  on this opening angle depends upon the reconstructed flight distance
  and \ip resolution of the two electrons.
\end{enumerate}

\begin{table}
  \begin{tabular}{l|cc|cc}
    \toprule
    requirement & 
    $S^\mathrm{(i)}_\TM$ & 
    $B^\mathrm{(i)}_\mathrm{tot}$ & 
    $S^\mathrm{(ii)}_\TM$ & 
    $B^\mathrm{(ii)}_\mathrm{tot}$ \\
    \midrule
    base 
    & $3.4\times10^3$ 
    & $3.2\times10^7$ 
    & $1.6\times10^3$
    & $5.4\times10^6$\\
    $\doca(\track, e)$ 
    & $3.0\times10^3$ 
    & $8.5\times10^6$ 
    & $1.3\times10^3$ 
    & $1.1\times10^6$ \\
    $\theta$ 
    & $1.5\times10^3$ 
    & $1.8\times10^4$ 
    & $6.4\times10^2$ 
    & $1.9\times10^3$ \\
    \midrule
    efficiency
    & $4.4\times10^{-1}$  
    & $5.6\times10^{-4}$ 
    & $4.0\times10^{-1}$  
    & $3.5\times10^{-4}$ \\    
    \bottomrule 
  \end{tabular}
  \caption{Expected signal and background yields for the
    $ee\,(ee\gamma)$ final state label as i\,(ii), assuming 100\,\%
    reconstruction efficiency for the final state and a collected Run
    3 dataset of $15~\ifb$.
    \label{tab:eeN}}
\end{table}

The numbers of expected \TM candidates are given in
Table~\ref{tab:eeN} for the signal and background after the baseline
selection, as well as after each of the two additional requirements.
Less than 0.1\,\% of the signal events pass the baseline selection,
largely due to the inefficiency of the $\pt$ requirements; however,
the $\pt$ selections cannot be significantly loosened. The
efficiencies of the additional selections beyond baseline, however,
are of order one for the signal and $\sim10^{-3}-10^{-4}$ for the
background, allowing for efficient background reduction. There is an
additional efficiency for reconstruction of all the final-state
particles, $\varepsilon_f$, which originates from the reconstruction
of the tracks, both online and offline, and from applying particle
identification criteria. Because the expected electron and photon
efficiencies are not yet public for Runs~3 and~4, we leave
$\varepsilon_f$ as an unspecified quantity in our expression for the
significance and discuss the implications shortly. We note that final
state reconstruction efficiencies can be estimated based on current
LHCb performance. From the $B\to J/\psi K^{*0}$
analysis~\cite{Nicol:2012ica} we find that
$\varepsilon_{e^+e^-}>10$\,\%, and from Ref.~\cite{LHCb:2012af} we
estimate $\varepsilon_{\gamma e^+e^-} \approx0.3\,\varepsilon_{e^+e^-}
> 3\,$\%. For further details see Appendix~\ref{sec:LHCbPref}.

Because the background rate in the signal region can be estimated
using the invariant mass sidebands, we expect the significance to be
limited by the statistical uncertainty of the sample. The LHCb
inclusive dark photon di-muon search~\cite{Aaij:2017rft} successfully
used such a technique~\cite{Williams:2017gwf}, although inclusion of
known background structure helped improve significance. The shape of
the $B$-hadron background has been demonstrated to be well
modeled~\cite{Aaij:2017rft}, and there is a similar expectation for
this analysis. Therefore, the \TM signal significance is approximately
given by

\begin{align}
  \label{eq:Sig}
  \sig \approx \frac{S_{\TM}}{\sqrt{B_\mathrm{tot}}}
  \sqrt{\frac{\varepsilon_f\cL}{15\,\ifb}} \, ,
\end{align}

where $S_\TM$ and $B_\mathrm{tot}$ are the expected number of signal
and background events from Table~\ref{tab:eeN}, $\varepsilon_f$ is the
final state reconstruction efficiency, and $\mathcal{L}$ is the
integrated luminosity of the dataset. Using the expected Run 3 dataset
of $15\,\ifb$, \TM can be discovered with $\sig \geq 5$ when
$\varepsilon_f > 20\,\%\,(12\,\%)$ for the $e^+ e^- (e^+e^-\gamma)$
final state. Given the current LHCb performance, these efficiencies
are realistic; see the above discussion and Appendix~\ref{sec:LHCbPref}.
In Fig.~\ref{fig:Sig} we plot the required integrated
luminosity for discovery of \TM, \textit{e.g.} $\sig \geq 5$, as a
function of $\varepsilon_f$.

\begin{figure}[t]
\includegraphics[width=0.5\textwidth]{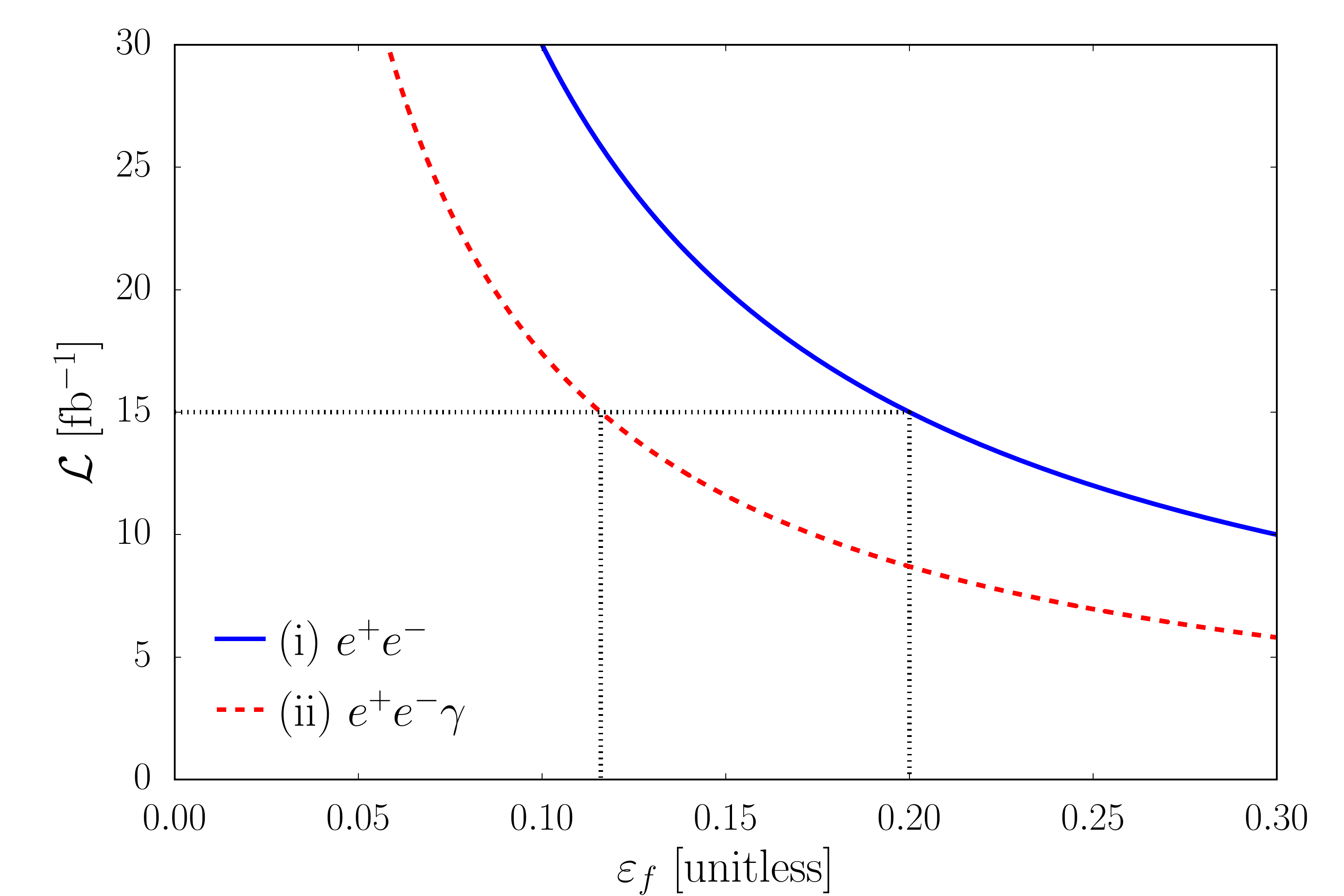}
\caption{ The required integrated luminosity for a $5\,\sig$
  discovery of \TM as function of the final reconstruction efficiency,
  $\varepsilon_f$ for the proposed (blue) $e^+ e^-$ and (red) $e^+ e^-
  \gamma$ searches.
  \label{fig:Sig}}
\end{figure}

In addition, Fig.~\ref{fig:dist} shows the differential cross sections
with respect to the $e^+e^-$ invariant mass for signal and
combinatorial background at LHCb, assuming a global efficiency to
reconstruct the \TM candidates of 20\%, or 6\% when also considering
the reconstruction of the additional photon from the $\eta$ decay.

\begin{figure*}[t]
  \includegraphics[width=0.48\textwidth]{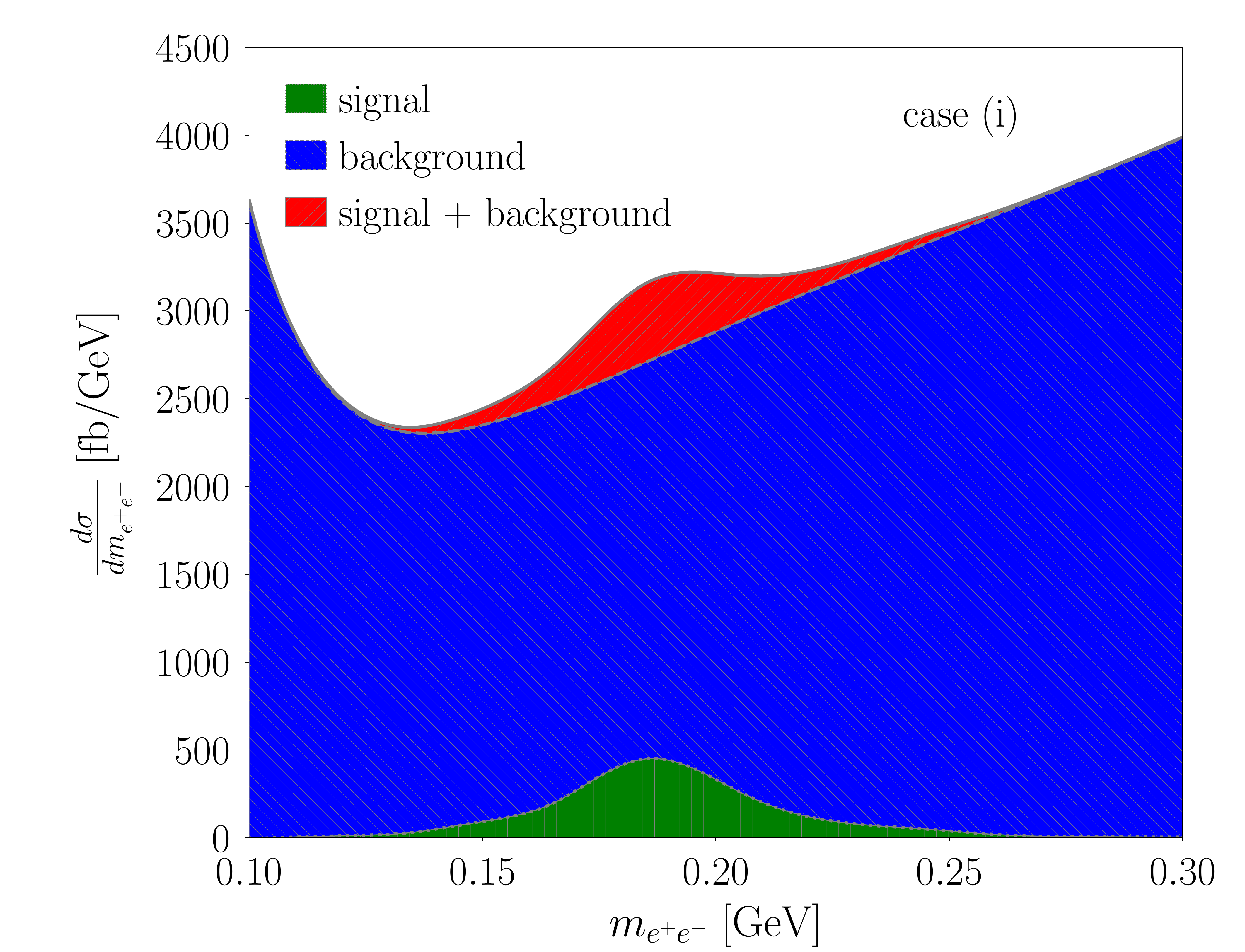} 
  \includegraphics[width=0.48\textwidth]{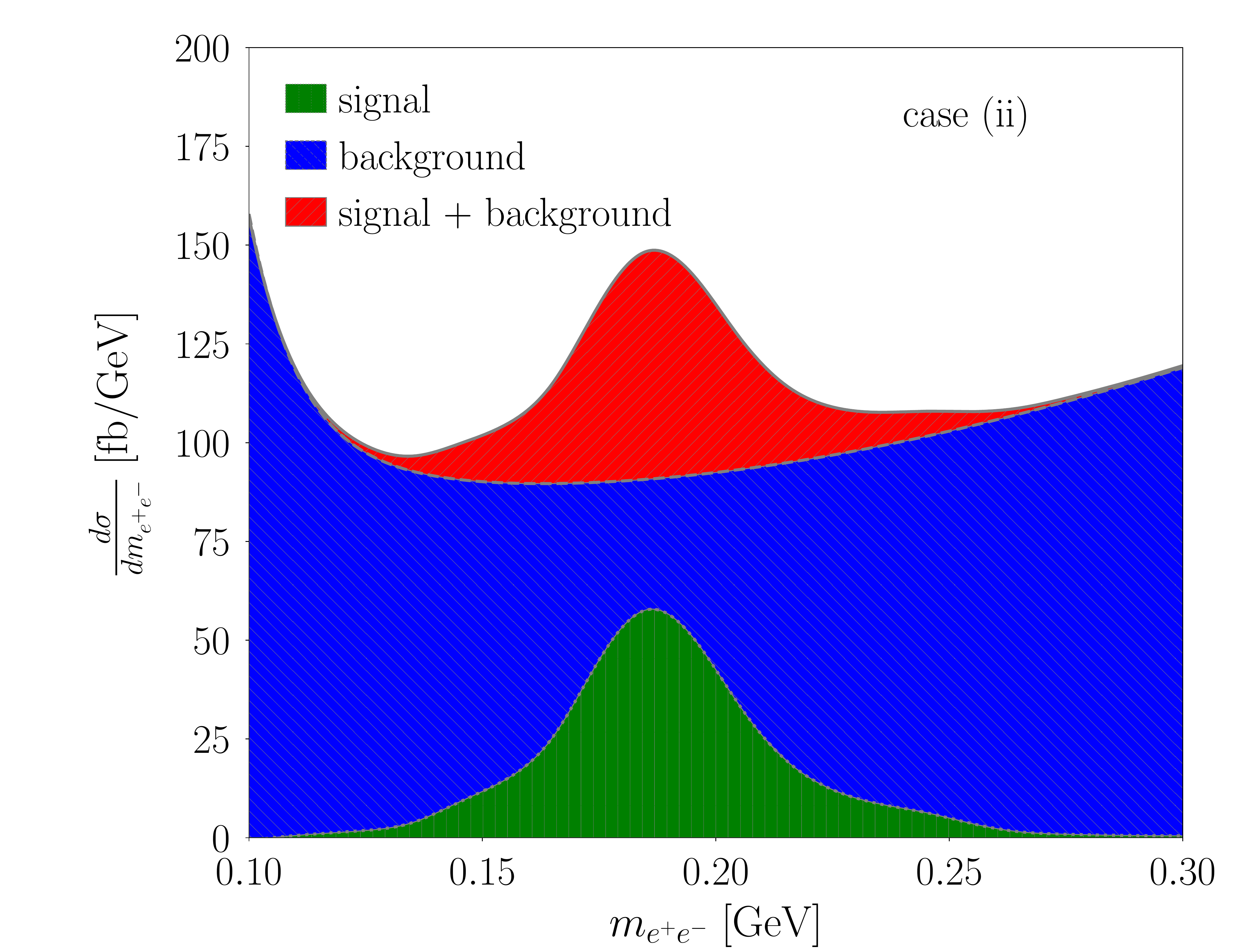} 
  \caption{The differential cross sections with respect to the
    $e^+e^-$ invariant mass for the expected \TM signal and
    combinatorial background at LHCb, assuming the normalization in
    Tab.~\ref{tab:eeN} for case (i) and (ii). Global efficiencies of
    20\% and 6\% are assumed to reconstruct the \TM and \TM plus
    photon candidates, respectively. In these conditions, a
    $5\,\sigma$ observation would be possible with an integrated
    luminosity of 15\,fb$^{-1}$ in case (i) and of 30\,fb$^{-1}$ in
    case (ii). The invariant mass resolution of the signal is
    described in the text. The shift observed in the central position
    of the signal peak, due to the lack of reconstructed
    bremsstrahlung from the electrons, is compatible with that of
    Ref.~\cite{MarinBenito:2193358}. For the combinatorial background,
    the resulting invariant mass distribution is obtained from
    simulation.
    \label{fig:dist}}
\end{figure*}

We conclude this section by commenting that we considered additional
selection criteria that we found to be sub-optimal and therefore did
not include in our analysis. First, we can require that the \ip of the
two electrons, projected onto the normal of the decay plane, is
consistent with zero. The decay plane is defined by the first hit of
each electron track and the primary vertex. We found that this
observable does not provide strong separation after the above
selection has been applied. Second, the expected proper lifetime of
the \TM candidate is known, and so in principle the transverse flight
distance can be used to select events that are most consistent with
this hypothesis. However, the analysis is more robust if no assumption
is placed on the lifetime of the \TM candidate, and it does not appear
to be necessary to reach a $5\,\sig$ discovery significance.

\section{Discussion and outlook}
\label{sec:outlook}
As outlined above, we project that LHCb will be able to discover $\TM$
with a statistical significance exceeding $5\,\sig$ in Run~3.
Ultimately, LHCb and other experiments can directly measure the $\TM$
mass, lifetime, and production rate (from $\eta$ decays or other
mechanisms). Since the \TM properties are well predicted by the SM,
this will be a test of the SM predictions in
Eqs.~\eqref{eq:mTM}--\eqref{eq:tauTM}, and any deviation from them is
a clear sign of new physics coupled to muons. Examples include dark
photons, $L_\mu-L_\tau$ gauge bosons, scalars, or axion-like
particles. In the presence of any of these particles, the \TM mass
(via the binding energy), lifetime, branching ratios and spectroscopy
(see discussion in~\cite{Jaeckel:2010xx}) are modified, and thus \TM
measurements can discover or constrain new muonic interactions. Such
new forces are motivated by several possible discrepancies with
predictions of the SM in other experiments, including measurements of
the muon anomalous magnetic moment $(g-2)_\mu$~\cite{Bennett:2006fi},
and the proton charge radius
problem~\cite{RevModPhys.88.035009,Beyer79,Carlson:2015jba,Epstein:2014zua}.
However, strong constraints on new physics exist from direct
searches~\cite{TheBABAR:2016rlg,Sirunyan:2018nnz}, measurements of
$(g-2)_\mu$, neutrino
experiments~\cite{Mishra:1991bv,Bellini:2011rx,Harnik:2012ni,
  Agostini:2017ixy,Altmannshofer:2014pba,Magill:2016hgc,
  Altmannshofer:2019zhy}, and $e\mu$
spectroscopy~\cite{Karshenboim:2010cg,Karshenboim:2014tka,Frugiuele:2019drl}.
Indeed, these constraints are generally more powerful than the
expected sensitivity of LHCb to $\TM$, although some exceptions exist
(for example, $(g-2)_\mu$ constraints can be alleviated if there are
other new particles whose effects partially cancel). New muonic forces
can also be probed as in
Refs.~\cite{Gninenko:2014pea,Grange:2015fou,Chen:2017awl,Abe:2019thb,
  Kahn:2018cqs,Krnjaic:2019rsv}. For a detailed analysis see
Appendix~\ref{sec:BSM}.

In the context of this study, we also considered the possibility of an
inclusive search for a $\tau^+\tau^-$ bound state, see \textit{e.g.}
Ref.~\cite{Fael:2018ktm}. In particular, ortho-tauonium, with a
significant branching fraction to $\mu^+\mu^-$, would appear to be the
best candidate for an LHCb search. We find, however, that the short
lifetime of the tauonium (close to the $\tau$ itself), and the small
signal yield compared to the background make the prospects very poor
for being observed at LHCb.

In summary, we have studied the potential for LHCb to discover an
as-yet-undiscovered long-lived particle in the SM:~the $\mu^+\mu^-$
true muonium bound state. We have proposed a search for the vector
$1^3S_1$ true muonium state, $\TM$, which kinetically mixes with the
photon and decays to $e^+e^-$. We have demonstrated that a search for
$\eta\rightarrow \gamma\TM,\,\TM\rightarrow e^+e^-$ can exceed a
$5\,\sig$ statistical significance using a displaced vertex search,
and we have presented two possible searches:~an inclusive search for
the $e^+e^-$ vertex, as well as an exclusive search where we
reconstruct the additional photon and require $m(\gamma,\TM) =
m_\eta$. Since $\TM$ mixes kinetically with the photon and has a
signature similar to the dark photon, this method could also have
sensitivity to dark photons in a similar mass window.

\begin{acknowledgments}
We thank Gil Paz, Mike Williams and Jure Zupan for useful discussions, and
Johannes Albrecht, Matthew John Charles, Maxim Pospelov, Mike Williams
and Jos\'e Zurita for comments on the manuscript. We also thank the
organizers of the ``New ideas in detecting long-lived particles at the
LHC'' workshop at LBL for a stimulating environment for discussions,
along with other members of our working group:~Jeff Asaf Dror, Maxim
Pospelov, Yuhsin Tsai and Jos\'e Zurita.

The work of XCV is supported by MINECO (Spain) through the Ram\'{o}n y
Cajal program RYC-2016-20073 and by XuntaGal under the ED431F 2018/01
project.  PI is supported by a Birmingham Fellowship.  JP is supported
by the UK Science and Technology Facilities Council.  The work of BS
is supported by the U.S.~National Science Foundation under Grant
PHY-1820770.
\end{acknowledgments}

\appendix
\section{Signal and Background Simulation}\label{app:simulation_details}
All signal and background samples are simulated using
\textsc{Pythia\,8.240}~\cite{Sjostrand:2014zea}. The signal from
$\eta$ meson decays is generated using the flag \texttt{SoftQCD:all =
  on}, while the $B$-hadron background is generated using the flag
\texttt{HardQCD:bbbar = on}. For the latter, the \texttt{HardQCD} flag
in conjunction with repeated $B$-hadron decays was used to generate a
sufficiently large background sample. The results from this large
sample were found to be in agreement with a smaller background sample
generated using the more inclusive \texttt{SoftQCD}
configuration. Additionally, including more sophisticated $B$-hadron
decays using \textsc{EvtGen}~\cite{Lange:2001uf} was found to have no
noticeable effect on the final result. This is because \textsc{Pythia}
already uses the branching fraction tables from \textsc{EvtGen}, and
many of the inclusive \textsc{EvtGen} decays use \textsc{Pythia} for
showering and hadronization. The results from \textsc{Pythia} for both
signal and background are demonstrated to be reliable, with the
\textsc{Pythia} study of Ref.~\cite{Ilten:2016tkc} accurately
predicting the reach of the LHCb inclusive $\mu^+ \mu^-$ dark photon
search~\cite{Aaij:2017rft}.

Conversion backgrounds were estimated using the photon flux generated
from \textsc{Pythia} configured with the flag \texttt{SoftQCD:all =
  on}, and modeling the expected conversion rate within the material
of the upgraded LHCb detector. The cross section for photon
conversions was calculated using a method~\cite{Davies:1954zz},
similar to that implemented in the material simulation package
\textsc{Geant}~\cite{Agostinelli:2002hh}. The approximation of the
opening angle between the converted electron-positron pair is
under-estimated at high masses by the \textsc{Geant}
model~\cite{Brun:1994aa}, and so a correction was applied to produce
an invariant mass spectrum of the converted pair that matches the full
analytic expression~\cite{Tsai:1973py}.

\raggedbottom

\section{LHCb Performance}
\label{sec:LHCbPref}

\subsection{Invariant mass resolution and reconstruction efficiencies}
An upgraded version of the LHCb detector will record the result of
proton-proton collisions at $\sqrt{s}=14$\,\TeV during Runs 3 and 4 of
the LHC. Similar, if not better, performances of the detector are
expected during that period~\cite{Bediaga:2012uyd}. The upgrade of the
detector is currently taking place. One important feature of this
upgrade is the expected triggerless readout~\cite{Bediaga:2014vzo},
removing the need for a first-level hardware trigger that is present
in other LHC detectors. This will allow a dramatic increase in the
efficiency to reconstruct low-momentum signatures, such as the decay
products of \TM.

\flushbottom

An estimation of the efficiency to reconstruct the \TM candidates can
be achieved by comparing to other LHCb analyses containing an $e^+e^-$
final state. In the $B^0 \rightarrow J/\psi K^{*0}$ analysis, with the
$J/\psi$ decaying to an $e^+e^-$ pair, reconstruction and selection
efficiencies at the level of 5\% could be achieved during the first
years of LHCb running~\cite{Nicol:2012ica}. This efficiency includes
the reconstruction of the accompanying $K^{*0}$ particles decaying to
$K\pi$ pairs as well as selection cuts on the mother $B$
candidate. The kinematics of the selected \TM signal electrons and
those from $J/\psi$ decay have been checked to be in reasonable
agreement. Therefore, reconstruction and selection efficiencies above
10\% should be easy to achieve. Since the performance of the upgraded
LHCb detector is still to be determined, we chose to show the expected
significance as a function of the final state reconstruction
efficiency, rather than choosing a fixed value. This efficiency will
also account for additional selection requirements to be applied in
the experimental analysis.  This includes the use of particle
identification cuts or more sophisticated variables to discriminate
against the combinatorial background. In the same regard, additional
potential inefficiencies in the online reconstruction at the upgraded
detector can be factorized as part of that efficiency. It should be
remarked that the 5\% efficiency, given as a baseline above, already
includes this online reconstruction in the current detector.

One of the main challenges to reconstruct low momentum electrons at
LHCb is the fact that the magnet sweeps away an important fraction of
these particles, which then only leave hits in the pre-magnet tracking
stations. Therefore, these electrons can be reconstructed, but their
momenta are unknown. However, for the reconstruction of the \TM mass,
the knowledge of the $pp$ collision vertex (where the \TM was
produced), the \TM decay position, and the directions and momenta of
the decay electrons is over-constrained. In this case, only the full
reconstruction of one of the final-state electrons is necessary. For
the other electron, only the direction is needed, such that hits in
the pre-magnet tracking stations would be sufficient. The use of this
technique could significantly increase the reconstruction efficiency
of the \TM final state. One drawback of reconstructing electrons that
are swept away by the magnet is the missing information from the PID
detectors located after the magnet, \textit{e.g.} RICH~2, the
calorimeters, and the muon system. However, the PID information from
RICH~1, specially designed for low-momentum
particles~\cite{Adinolfi:2012qfa}, would still be available.
 
Concerning the $e^+e^-$ invariant mass resolution for the \TM
reconstruction, Ref.~\cite{MarinBenito:2193358} claims an invariant
mass resolution of $\sim8$\% to reconstruct $K_S^0 \rightarrow
e^+e^-e^+e^-$ decays at LHCb. The kinematic cuts in that study are
softer with respect to this one, and therefore the momentum resolution
for the electrons in this analysis is expected to be better, due to
the smaller effect of multiple scattering. However, here we assume a
similar invariant mass resolution, taking $\res\sim20\,\MeV$ with
radiative tails based on the invariant mass distribution from $K_S^0
\rightarrow e^+e^-e^+e^-$ decays. This conservative approach can be
confirmed by the $\res$ distribution from $B^0 \to J/\psi K^{*0}$
decays, with the $J/\psi$ decaying to an $e^+e^-$
pair~\cite{Aaij:2017vbb}. For these decays, using final state
electrons in a kinematic range similar to this study, resolutions at
the level of 2\% can be achieved with LHCb. The kinematic constraint
mentioned above, arising from the knowledge of the \TM decay position
and the $pp$ collision point, could also be used to improve the
$e^+e^-$ invariant mass resolution by $\approx 20\%$.

The full reconstruction of the $\eta \rightarrow \gamma \TM $ decay
also requires the determination of the reconstruction efficiency of
the $\gamma$. To obtain this, Ref.~\cite{LHCb:2012af} is used,
aligning our $\gamma$ selection cuts with those in that analysis. In
that study, an efficiency of 10\% is claimed to reconstruct the
photon. This includes both the effect of the kinematic cuts applied
and of the reconstruction in the LHCb ECAL. If the effect of the
kinematic cuts is factored out, an efficiency of $\approx 30$\% is
obtained.  This is taken as a baseline for this analysis. In order to
estimate the $\eta \rightarrow\gamma \TM$ decay invariant mass
resolution, an estimate of the $\gamma$ momentum resolution is
needed. This has two components, the direction and energy resolution
of the photons. The first depends on the ECAL cell size and on its
distance to the $pp$ collision point. Most of the signal photons are
found to fall in the most inner region of the ECAL, where the cells
have a size of $\approx 4\,\cm$~\cite{Aaij:2014jba}. This provides an
angular resolution of $\approx 0.002$. For the energy resolution,
Ref.~\cite{Aaij:2014jba} reports $\delta E/E \simeq 9\% \sqrt{\GeV/E}
\oplus 0.8\%$. Combining both effects together, an invariant mass
resolution of $\resg\approx 50\,\MeV$ is obtained. The methodology is
validated using multiple LHCb analyses with $\gamma$ in the final
states~\cite{LHCb:2012af,Aaij:2016ofv}.

\subsection{Impact parameter and DOCA resolution}
The description of the upgraded LHCb vertex locator (VELO) is taken
from Ref.~\cite{Bediaga:2013bkh}, using a nominal single hit
resolution of $12\,\mu\mathrm{m}$ in $x$ and $y$. Multiple scattering
is modeled~\cite{Lynch:1990sq} assuming an RF-foil thickness of
$0.25\,\mm$ and sensor thicknesses of $0.2\,\mm$. This material
description is validated against the full LHCb upgrade simulation
where the transverse impact parameter for a track is
parameterized by,

\begin{align}
  \sigma_{\ip} = \left(1.1 + \frac{1.3\,\GeV}{\pt}\right)\times10^{-2} \,
  \mm \, ,
\end{align}

where the first term is determined by the detector geometry and the
second term arises from multiple scattering. The uncertainty on the
distance of closest approach~(\doca) between two tracks is well
approximated as,
\begin{align}
  \sigma_\doca = \sigma_\ip^{(1)} \oplus \sigma_\ip^{(2)} \, ,
\end{align}
given $\sigma_\ip^{(1)}$ and $\sigma_\ip^{(2)}$ are the \ip
uncertainties for the first and second track, respectively.

\section{Muonium and Physics Beyond the Standard Model}  
\label{sec:BSM}
Since the properties of \TM are completely determined by the SM, the
ability of LHCb to independently measure the mass, production rate,
and lifetime of \TM provides the possibility of a precision test of
the SM. New particles and forces coupled to muons, including dark
photons, $L_\mu-L_\tau$ gauge bosons, low-mass scalars, and axion-like
particles, could potentially alter the muon binding energy and \TM
decay rates by providing additional annihilation channels for the
$\mu^+\mu^-$ bound state. Such new muonic forces have already been
predicted in the context of the persistent anomalous measurements of
$(g-2)_\mu$ and the proton charge radius problem, see \textit{e.g.}~\cite{TuckerSmith:2010ra}.

Here, we focus on BSM contributions to the \TM decay rate, both to SM
states mediated by new interactions but also the \TM decay to
hidden-sector states. Since the \TM production rate depends on the
\TM wavefunction at the origin, a new force can only appreciably
modify this if its structure constant is comparable to
$\alpha$. However, this structure constant is strongly constrained by
$(g-2)_\mu$ and other precision measurements. Therefore, the prospects
for BSM modifications to the \TM decay are more promising than for its
production, although still challenging to observe.

\subsection{Hidden-Sector Models} 
\label{sec:models}
We consider the following scenarios, which give rise to modifications
of the \TM decay rate and branching fractions:

\begin{align}
  	\text{Scalar ($S$): } 
	\cL_S= &y_{S\mu}\, S\bar\mu \mu+y_{Se}\,S\bar e e \, , \\
  	\text{Pseudoscalar ($a$): } 
	\cL_a = &y_{a\mu}\,a\bar\mu\gamma^5\mu + y_{a e}\,a\bar e\gamma^5 e
        \nonumber \\
	&+ \frac{g_{a\gamma}}{4}\,aF_{\mu\nu}\tilde{F}^{\mu\nu} \, , \\
	  \text{Vector ($V$): } 
	  \cL_V =& g_{V\mu}\, \bar \mu\gamma^\nu \mu V_\nu  \nonumber \\
	  &+ g_{Ve} \, \bar e\gamma^\nu e V_\nu \, , \\
	  \text{ Axial Vector ($A$): } 
	  \cL_A = &g_{A\mu}\, \bar\mu\gamma^\nu\gamma^5 \mu A_\nu  \nonumber \\
	  &+ g_{Ae} \, \bar e\gamma^\nu\gamma^5 e A_\nu \, ,
\end{align}

where $\tilde{F}^{\mu\nu} = \varepsilon^{\mu\nu\rho\sigma}
F_{\rho\sigma}/2$.

\subsection{\TM decay to a photon and a mediator} 
\label{sec:mediatordecay}
If the mediator $X=S,a,V$ or $A$ couples to muons, we can have decays
as $\TM\rightarrow \gamma X$ or $\TM\rightarrow XX$. Since the decay
to two mediators is typically suppressed by the square of the mediator
coupling to muons, the decay to $\gamma X$ is the most important.
Depending on the lifetime of $X$ and its decay modes, the signature
can be mono-photon, or photon and $e^+e^-$. Assuming that $\Gamma_\TM
\approx \Gamma_{\TM\to e^+e^-}$, see Eq.~\eqref{eq:tauTM}, we find the
following branching ratios

\begin{alignat}{2}
  \label{eq:BRgammaS}
  \br(\TM\rightarrow \gamma S) =& \frac{ y_{S\mu}^2}{2\pi\alpha
    (1-x_S)} \left( 1+4x_S+x_S^2\right) \, , \\
  \br(\TM\rightarrow \gamma a) =& \frac{(1-x_a)}{2\pi\alpha} 
  	\bigg[ y_{a\mu}^2  \nonumber \\
	& \qquad +g_{a\gamma}^2 m_{\TM}^2 \frac{(1-x_a)^2}{16} \bigg] \, , \\
  \br(\TM\rightarrow \gamma V) = &0 \, , \\
  \label{eq:BRgammaA}
  \br(\TM\rightarrow \gamma A) = &\frac{g_{A\mu}^2 }{2\pi\alpha}
  \frac{1+ 10 x_A + x_A^2}{1-x_A} \, ,
\end{alignat}

where $x_X = m_X^2/m_\TM^2$ and we neglect the relative momentum of
the muons in the \TM state. This is reasonable because this kinetic
energy is a small contribution to the energy released in the \TM
decay.

The limits on the coupling of the mediator to muons is generally model
dependent. However, the measurement of $(g-2)_\mu$ provides a
sensitive probe of new physics coupled to muons. In principle, it is
generally possible to evade these constraints by having another
contribution to $(g-2)_\mu$ that almost cancels the one from the
mediator. In Fig.~\ref{fig:MaxBR} we plot the maximal \TM branching
ratio to final states in Eqs.~\eqref{eq:BRgammaS}--\eqref{eq:BRgammaA}
which is allowed by measurements of $(g-2)_\mu$ at the $5\,\sigma$
level, \textit{i.e.} $\Delta a_\mu =
\frac{1}{2}(g-2)_{\mu}(\mathrm{obs}) -
\frac{1}{2}(g-2)_{\mu}(\mathrm{SM}) \in \left[-1.1, 6.9\right] \times
10^{-9}$~\cite{Bennett:2006fi}. We do not include the effects of the
coupling $g_{a\gamma}$ on the branching fraction to pseudoscalars
because of the powerful constraints on direct searches for axion-like
particles from LEP data, which lead to a negligible contribution to
the \TM branching fraction into
pseudoscalars~\cite{Knapen:2016moh,Jaeckel:2015jla,Kahn:2016vjr}; see
also a recent recast of PrimEx data~\cite{Larin:2010kq,Aloni:2019ruo}.
The expressions for NP contributions to $\Delta a_\mu$ are taken
from~\cite{Batell:2016ove,Chen:2015vqy,Kahn:2016vjr}. As we can see
the maximal branching ratios are typically below the 1\,\% level and
require high precision \TM measurements to exceed this sensitivity.

\begin{figure}
  \includegraphics[width=0.49\textwidth]{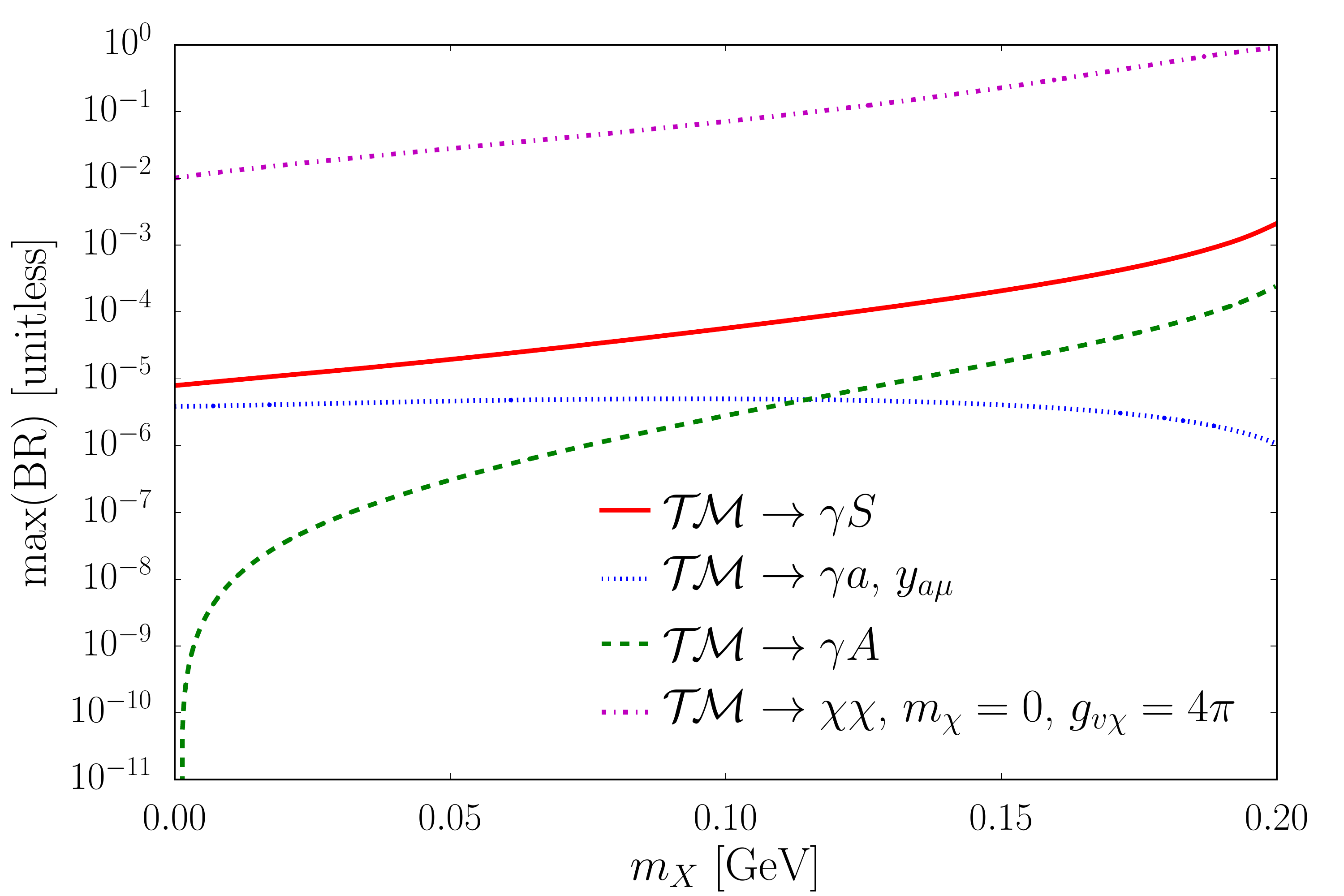} 
  \caption{\TM branching ratio to BSM final states in
    Eqs.~\eqref{eq:BRgammaS}--\eqref{eq:BRgammaA} which  are allowed by
    $(g-2)_\mu$ at the $5\,\sigma$.
    \label{fig:MaxBR}}
\end{figure}

\subsection{\TM decay to hidden-sector particles} 
\label{sec:HSdecay}
In this section, we calculate decay rates of \TM to hidden-sector
particles $\chi$ such as $\TM\rightarrow\bar\chi\chi$, where $\chi$ is
a hidden-sector particle. This results from muon annihilation via an
$s$-channel mediator into the $\chi$ particles. This final state
dominates when the mediator has a much larger coupling to hidden
sector particles than SM particles. These $\chi$ particles could be
invisible, or in turn decay to lighter hidden-sector particles. We
consider the same mediators as in Section~\ref{sec:mediatordecay} and
assume that $m_\TM\neq m_X$; otherwise, we have to take into account
mixing between the states. We note that the SM rate of $\TM \to Z^*
\to \bar\nu\nu$ is completely negligible.

Let us assume for concreteness that $\chi$ is a Dirac fermion. The
coupling to $\chi$ has the same parity structure as to SM leptons,
\textit{e.g.} we assume that a scalar couples to $\bar\chi\chi$, a
pseudoscalar to $\bar\chi\gamma^5\chi$, etc. Because the \TM state we
are considering is a vector, the only contribution is via decay
through a vector state. Then, we have

\begin{align}
  \br(\TM\rightarrow V^*\rightarrow\bar\chi\chi)
  = & \frac{g_{V\mu}^2 g_{V\chi}^2}{16\pi^2\alpha^2 (1-x_V)^2}
  \nonumber \\
  & \times \left(1+2x_\chi \right)\sqrt{1-4x_\chi} \, .
\end{align}

If we consider $m_V < m_{\TM}$ such that there is no suppression of
the $V$ propagator and $m_\chi \ll m_{\TM}$, we obtain constraints on
the coupling $g_{V\mu}$ from $(g-2)_\mu$. The coupling to $\nu_\mu$
leads to constraints on neutrino trident rates, so for a vector
coupling these also constrain $g_{V\mu}\,$. The maximal allowed value
of $\br(\TM\rightarrow V^*\rightarrow\bar\chi\chi)$ by $(g-2)_\mu$ for
$g_{V\chi}=4\pi$ and $m_\chi=0$ is plotted in Fig.~\ref{fig:MaxBR}.
For $m_V\ll m_{\TM}$, this gives a hidden-sector branching fraction at
the level of 2\%. While this is likely too small to be seen as a
change in the \TM lifetime or cross section, it could be detectable if
the $\chi$ decays themselves are visible, which is challenging. If
$m_V=160\,\MeV$, the branching fraction is enhanced to $\sim 10\%$. If
the states become much more degenerate than this, it is: (a) tuned;
(b) would require some careful treatment of the width and mixing
between the two states. This is especially true if the coupling
$g_{V\chi}$ is very large, because the width would be large as
well. If we instead take $g_{V\chi}=1$, then the branching fraction is
$\sim10^{-4}$ for $m_V=0$ and $\sim10^{-3}$ for $m_V=170\,\MeV$.

\subsection{Modifications to \TM decay to $e^+e^-$}
\label{sec:epemdecay}
In this section, we consider $s$-channel contributions of the mediator
to the decay of $\TM\rightarrow e^+e^-$. This is similar to the decay
from Section \ref{sec:HSdecay}, but we must include interference with
the contribution from the SM photon. We obtain (in the limit $m_e \ll
m_{\TM}, m_V$)

\begin{align}
  \Gamma(\TM\rightarrow e^+e^-) =& \frac{\alpha^3}{192\pi^2 (1-x_V)^2}
  \nonumber \\
  & \times \left[g_{V\mu}g_{Ve}+4\pi\alpha(1-x_V)\right]^2 \, ,
\end{align}

which appropriately reduces to the $V$-only or photon-only results in
the limits $\alpha\rightarrow0$ and $g_{V\mu}=g_{Ve}=0$, respectively.
For $g_{V\mu},\,g_{Ve}\ll\sqrt{4\pi\alpha}$, the dominant correction
to the width from the SM value scales like

\begin{align}
  \frac{\Delta \Gamma}{\Gamma} =
  \left. \frac{g_{V\mu}g_{Ve}}{2\pi\alpha(1-x_V )} \right|_{x_V \ll 1}
  \lesssim 2\times 10^{-5}.
\end{align}

Note that we can apply this to a dark photon by simply choosing
$g_{Ve}=g_{V\mu}=\varepsilon\sqrt{4\pi\alpha}$, where $\varepsilon$ is
the kinetic mixing of the dark photon.

\twocolumngrid
\vspace{-8pt}
\section*{References}
\vspace{-10pt}
\def\bibsection{}
\bibliographystyle{utphys}
\bibliography{tm}

\end{document}